\documentclass[12pt]{iopart}
\usepackage{graphicx}
\usepackage{iopams}

\begin{document}

\title{Resonant tunneling through a C$_{60}$ molecular junction in liquid environment}

\author{Lucia Gr{\"u}ter$^1$, Fuyong Cheng$^2$, Tero T. Heikkil{\"a}$^1$, M. Teresa
Gonz\'{a}lez$^1$, Fran\c{c}ois Diederich$^2$, Christian
Sch{\"o}nenberger$^1$ and Michel Calame$^1$}

\address{$^1$Institut f{\"u}r
Physik, Universit{\"a}t Basel, Klingelbergstr.~82, CH-4056 Basel,
Switzerland }

\address{$^2$Laboratorium f{\"u}r Organische Chemie,
ETH-H{\"o}nggerberg, HCI, CH-8092 Z{\"u}rich, Switzerland}

\ead{michel.calame@unibas.ch}

\date{\today}

\begin{abstract}

We present electronic transport measurements through thiolated
C$_{60}$ molecules in liquid environment. The molecules were
placed within a mechanically controllable break junction using a
single anchoring group per molecule. When varying the electrode
separation of the C$_{60}$-modified junctions, we observed a peak
in the conductance traces. The shape of the curves is strongly
influenced by the environment of the junction as shown by
measurements in two distinct solvents. In the framework of a
simple resonant tunneling model, we can extract the electronic
tunneling rates governing the transport properties of the
junctions.

\end{abstract}


\maketitle


\section{Introduction}

How can we experimentally determine the relevant energy levels and
the tunneling rates involved in electronic transport through
single molecule devices? Such questions are of fundamental
interest for the development of molecular electronics. The
conductance of single molecules has been investigated in several
experiments where phenomena such as Coulomb blockade and Kondo
effect \cite{mceuen02, park02}, negative differential resistance
\cite{tour99} and logic gates \cite{heath99} were pointed out.
Different techniques were used to address single or a few
molecules. Fixed contact arrangements such as nanopore systems
\cite{reed97} and nanogaps fabricated by shadow evaporation or
electromigration \cite{mceuen99} have been realized.
Alternatively, scanning tunneling microscope (STM) \cite{cui01,
gimzewski99} and break junction techniques \cite{reed97,
Kergueris99, mayor02} offer tunable contacts to characterize
molecules. Break junctions offer an additional stability when
compared to the STM, but the implementation of a third electrode
remains delicate although recent work shows it can be done
\cite{ralph04}. Adding a chemically controllable environment in
break junctions experiments would provide an additional control of
the anchoring of the molecule to the metallic constriction and
allow \emph{in situ} electrochemistry experiments. The excellent
efficiency of a liquid gate bas been demonstrated previously on
carbon nanotube field-effect transistors \cite{krueger01}, and
recently this effect has been studied for organic molecules using
a STM~\cite{tao04,jackel04}.

Buckminsterfullerenes and in particular C$_{60}$ and its
derivatives have attracted much attention \cite{dresselhaus96_12}
and, due to their particular electronic properties, represent
ideal model systems for molecular electronics
\cite{gim97,park00,porath97}. In this Letter, we present
electronic transport measurements through thiolated C$_{60}$
derivatives in a two probe configuration provided by the tips of a
break junction (Fig.~\ref{figure1}). The single functional group
serves as an anchor and, in the case where a single molecule
junction is realized, permits to control the coupling of the
C$_{60}$ to a Au electrode by mechanically adjusting the
inter-electrode spacing $d$. By performing the transport
measurements in two distinct solvents, dimethyl sulfoxide (DMSO)
and toluene, we could observe the strong influence of the
environment on the molecular junctions.

\section{Experimental section}

The synthesis of the fullerene derivative was carried out
following well-established protocols \cite{shi94}. A thiolated C4
linker ensured that the C$_{60}$ molecule would bind to the gold
electrodes. To contact the molecules and perform transport
measurements, we used a mechanically controllable break junction
setup \cite{ruitenbeek96}. A gold nanobridge was lithographically
patterned on a phosphor bronze substrate. By bending the substrate
via a vertically moving rod, the gold bridge is elongated and
finally breaks. The two resulting ends serve as electrodes for
contacting the molecules. The setup includes a cell that allows
working in different liquid environments and has been described
previously \cite{grueter04}.

We first characterized the gold junctions via measurements of the
conductance $G(d)$ versus distance $d$ in air and in the solvents
used subsequently for the C$_{60}$ molecules, toluene and dimethyl
sulfoxide (DMSO). Several sets of consecutive open-close cycles
were recorded, with $G$ ranging typically between
\mbox{$0.001$\,$G_0$} and a few $G_{0}$, where $G_0=2e^2/h$ is the
conductance quantum. To investigate the transport through C$_{60}$
molecules, a \mbox{$0.1$\,mM} C$_{60}$ solution was added to the
liquid cell while the junction was kept close. The junction was
then opened widely \mbox{($\approx 5$\,nm)} to favor the
self-assembly of the molecules. We also added molecules while
keeping the junction open but no significant differences in the
transport measurements were observed. Sets of $10-30$ $G(d)$
curves were recorded at time intervals of \mbox{$30-45$\,min}.
During these cycles, the junction was never fully closed, keeping
its conductance below \mbox{$0.1$\,$G_0$} in order to limit
mechanical rearrangements of the Au tips.
$G(d)$-values obtained while closing the junction are shown in
Fig.~\ref{figure2} for DMSO (a) and toluene (b) at a bias
voltage of \mbox{$V_b=0.2$\,V}.

The open symbol curves show measurements in the pure solvent. The
solid curves are five successive raw measurements performed after
adding the C$_{60}$ solution to the junction. We also attempted to
measure current-voltage ($I-V$) characteristics. However, these
curves were not reproducible, displaying rather large uncontrolled
hysteresis, which we assign to the dynamics of the liquid
environment and the relatively weak tethering of the C$_{60}$ with
one single linker group only.

While the measurements of $G(d)$ in the pure solvents show a clean
exponential behavior \cite{grueter04}, the curves for the C$_{60}$
modified junctions exhibit a more complex structure. In previous
break-junction work with molecules, the occurrence of a plateau in
$G(d)$ has been taken as the signature for the `locking-in'
of a molecule~\cite{mayor02}. We observe a plateau in about $50\,\%$
of the measured closing cycles. In addition to this observation, a
pronounced peak can appear (arrows in Fig.~\ref{figure2}).
Though, the peak appears less frequently
in different samples, reproducible measurements
(visible in each closing curve within a cycle of $10$ curves at least) could be
acquired for both C$_{60}$ in DMSO and toluene. In DMSO, the peak
corresponds to a conductance maximum of \mbox{$\simeq 0.012$\,$G_{0}$}
and is followed by a plateau-like feature.
When further closing the junction the
conductance rises more rapidly, although less sharply than in the
pure solvent. The shape of the conductance curves in toluene
(Fig.~\ref{figure2}b) is very different, showing more noise and a
fully developed peak. After a small pre-peak (not present in all
curves), the conductance rises to a maximum value of $\approx
0.025 G_{0}$ at the peak and decays again to a very low
conductance. While further closing the junction the conductance
increases sharply, with a slope similar to that of the conductance
in the pure solvent. The environment in which the junction is
studied plays clearly an important role here. We note in
particular that DMSO is a rather poor solvent for neutral
fullerenes whereas toluene is a good one (see e.g.
\cite{marcus01}).

To calibrate each junction, we measured the tunneling conductance
between the two Au electrodes in the pure solvent. Let us denote the
gap distance between the two Au electrodes by $d_{gap}$.
It can be expressed as $d_{gap}=rs$, where $s$ denotes the vertical
displacement of the pushing rod and $r$ the reduction factor
which depends on the geometry of the junction \cite{ruitenbeek96}.
It is the reduction factor $r$ which needs to be calibrated for each junction.
The tunneling conductance $G(s)$ measured in a
pure solvent was fitted according to the expression
$G(s) \propto \exp[-\kappa r s]$ with $\kappa=2 \sqrt{2 m \phi}/\hbar$,
where $\phi$ (and $\kappa$) represents the height of a square tunneling barrier
of width $d_{gap}$ and $m$ is the electron mass. The fits were
adjusted by tuning $r$ in order to obtain values for $\phi$ in
agreement with our previous work \cite{grueter04} in which
the apparent workfunctions in different solvents
were calibrated with respect to measurements in vacuum.
The reference inverse decay lengths $\kappa$, valid for the pure
solvents, are \mbox{$\kappa_{toluene}=0.85$\,\AA$^{-1}$} and
\mbox{$\kappa_{DMSO}=0.63$\,\AA$^{-1}$}. This procedure provides
the reduction factor $r$ for each junction measured.
\cite{cui02}.

\section{Discussion}

\begin{figure}
\includegraphics*[width=10cm]{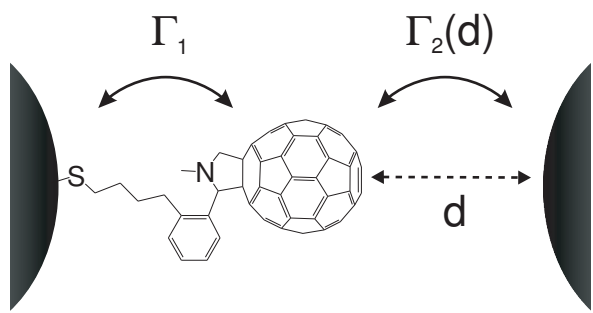}
\caption{\label{figure1} Schematic representation of a break
junction with a thiolated C$_{60}$ molecule anchored to the left
electrode. The distance $d$ between the molecule and the right
electrode can be adjusted by opening and closing the junction. }
\end{figure}

The particularity of our system lies in the single anchor group
used to tether the molecule within the break junction. This
configuration makes it possible to tune the coupling of the
molecule to the electrode: while one tunneling rate
($\Gamma_{1}/\hbar$) is fixed, the second ($\Gamma_{2}/\hbar$)
varies exponentially with the gap between the electrode (electrode to the right in Fig.~\ref{figure1})
and the C$_{60}$ molecule.
The metal--molecule--metal junction forms a double-barrier junction.
In this picture, the observed conductance
peaks can be explained within a coherent resonant tunneling model,
provided that one of the energy levels of the molecule lies
not too far away from the Fermi level of the electrodes (Fig.~\ref{figure3}a).

Resonant tunneling through a single level $\varepsilon$
results in a peak in the linear conductance $G$ when
the level aligns with the Fermi level $E_F$ of the leads, i.e.
at $\varepsilon=0$. The occurrence of a peak in $G(\varepsilon)$ is a well known
fact \cite{datta}. However, a peak can also arise
in $G(\Gamma_2)$, even if $\varepsilon \not= 0$.
Resonant tunneling is described by the Breit-Wigner
equation:

\begin{equation}\label{eq1}
G=\frac{2e^2}{h}\frac{\Gamma_1 \cdot
\Gamma_2}{\varepsilon^2+(\Gamma_1+\Gamma_2)^2/4}
\end{equation}

In our geometry, $\Gamma_{1}$ is fixed and defined by the molecular tether
holding the molecule to the left electrode (Fig.~\ref{figure1}).
$\Gamma_2$ depends exponentially on the distance $d$ between the
C$_{60}$ molecule and the right electrode with $\kappa$ being the
inverse decay length, i.e. $\Gamma_{2}(d)=\Gamma_{2}^{\ast}\cdot
e^{-\kappa d}$. $G(\Gamma_2)$ in Eq.~(\ref{eq1}) as a function of
$\Gamma_2$ has a maximum $G_{max}$ at
$\Gamma_2^{\ast}=\sqrt{4\varepsilon^2+\Gamma_1^2}$. For
$\varepsilon \ll \Gamma_1$ we would have
$\Gamma_2^{\ast}=\Gamma_1$ and consequently a peak conductance of
$G_{max} \simeq G_0$. In the experiment, the peak conductance is
however much smaller than $G_0$, so that $\Gamma_1 \ll
\varepsilon$. It follows then that $\Gamma_2^{\ast}\simeq
2\varepsilon$ and $G_{max}=G_0 \cdot \Gamma_1/\varepsilon$. A
respective plot in this limit ($\Gamma_1 \ll \varepsilon$) is
shown in Fig.~\ref{figure3}b. Note, that we define $\varepsilon
\geq 0$, but the sign of $\varepsilon$ is not determined, i.e. a
sign change in $\varepsilon$ does not affect the discussion.

Using the Breit-Wigner equation we can obtain the ratio $\Gamma_1/\varepsilon$
from the measured peak height without fitting.
The measurements yield $\Gamma_1/\varepsilon \simeq 1.1\cdot10^{-2}$ in DMSO and
$\Gamma_1/\varepsilon \simeq 2.7\cdot10^{-2}$ in toluene, more
than twice the DMSO value.
Because we expect $\Gamma_1$ to be the same for both solvents (after all it is the same linker),
this suggests that the level is lying closer to the Fermi level of the electrodes in
toluene as compared to DMSO.

That we can measure directly the ratio $\Gamma_1/\varepsilon$ is a
very nice fact. But we would like to obtain both $\varepsilon$ and
$\Gamma_1$. This is at first sight not possible, because
Eq.~(\ref{eq1}) is scale-invariant. Changing $\Gamma_1$,
$\Gamma_2$, and $\varepsilon$ by the same factor leaves the curve
invariant. To our rescue, we emphasize that we did not measure the
linear-response conductance, but $G$ at a bias of
\mbox{$V_b=0.2$\,eV}, much larger than $k_B T$ at room
temperature. The finite bias introduces an energy scale which
allows a considerable narrowing down of the range of possible
values.

Integrating Eq.~(\ref{eq1}) over the bias window, given by the
applied voltage $V_b$, yields:
\begin{equation}\label{eq2}
G={\frac{4e^2}{h}}{\frac{\gamma_1~\gamma_2(d)}{\gamma_1+\gamma_2(d)}}\left[\arctan\left(\frac{1-2\tilde{\varepsilon}}{\gamma_1+\gamma_2(d)}\right)+\arctan\left(\frac{1+2\tilde{\varepsilon}}{\gamma_1+\gamma_2(d)}\right)\right]
\end{equation}
All parameters are now scaled to $V_b$, i.e.
$\gamma_i=\Gamma_i/eV_b$, $\gamma_2(d)=\gamma_2^{\ast} \cdot
e^{-\kappa d}$, and $\tilde{\varepsilon}=\varepsilon /eV_b$. There
are two regimes: a) $\varepsilon \lesssim V_b/2$ and b)
$\varepsilon \gg V_b/2$, the latter corresponds to the
Breit-Wigner equation. In both cases a peak develops in
$G(\gamma_2)$, but there is a pronounced crossover, which sets in
sharply at $\tilde{\varepsilon}:=\varepsilon/eV_b=0.5$. This
crossover is shown in Fig.~\ref{figure3}c for a selected value of
$\gamma_1=0.02$. In the following, we will try to fit our
measurements to Eq.~(\ref{eq2}) in both the low and large
$\varepsilon$ limit. There are four fitting parameters:
$\varepsilon$, $\Gamma_1$, $\Gamma_2^{\ast}$, and $\kappa$. Note,
that we always use the ansatz $\Gamma_2=\Gamma_2^{\ast} \cdot
e^{-\kappa d}$ and that we define the zero of the distance axis
$d$ to match with the peak position in the measurement. This is
possible, as we have no means to determine the `true' zero, i.e.
the point of contact of the right electrode with the C$_{60}$
molecule.

\begin{figure}
\includegraphics*[width=11cm]{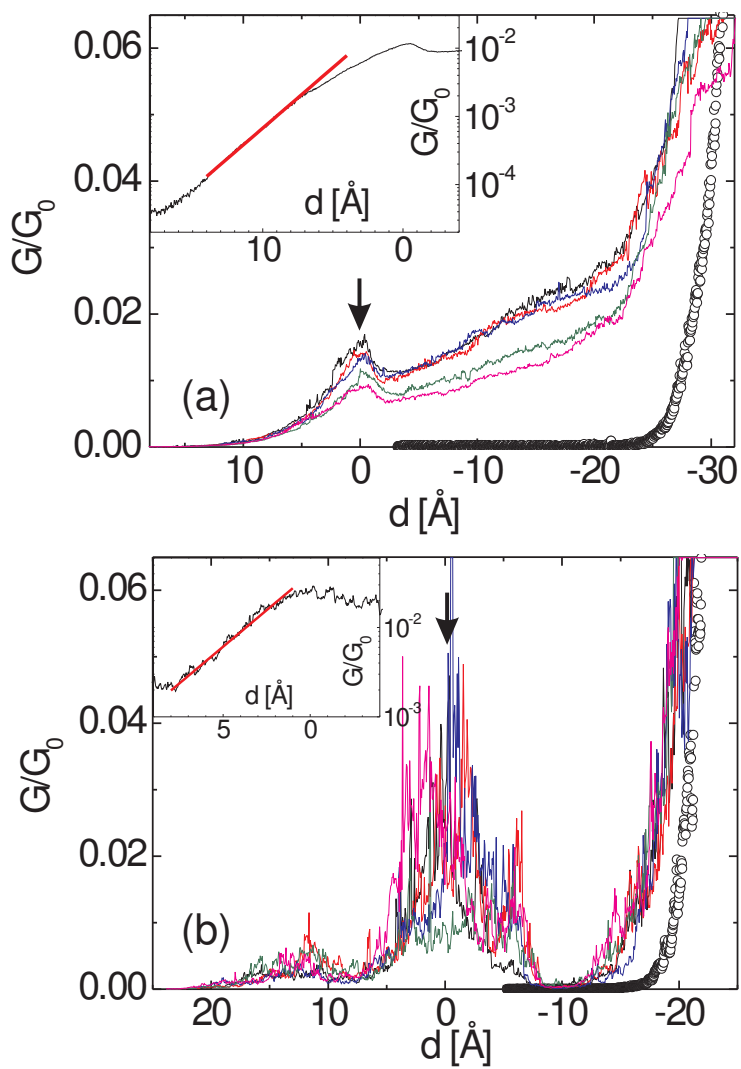}
\caption{\label{figure2} Conductance $G$ versus distance $d$ curves in
    DMSO (a) and toluene (b) measured at a bias voltage of \mbox{$0.2$\,V} while
    closing the junction. The open symbol curves were measured in the
    pure solvent. The solid lines are raw data for five successive
    measurements after adding the C$_{60}$ solution. The insets show
    averaged (over 10 curves) conductance data in a log-lin
    representation. The value for the inverse decay rate $\kappa$ was obtained from a linear
    fit (solid lines) to $G/G_0$ in the low conductance regime:
    $\kappa_{DMSO}=0.43$\AA$^{-1}$ and $\kappa_{toluene}=0.37$\AA$^{-1}$.
     Note, that the zero in the distance scale $d$ was chosen to match
     with the position of the peak.
    }
\end{figure}

\begin{figure}
\includegraphics*[width=11cm]{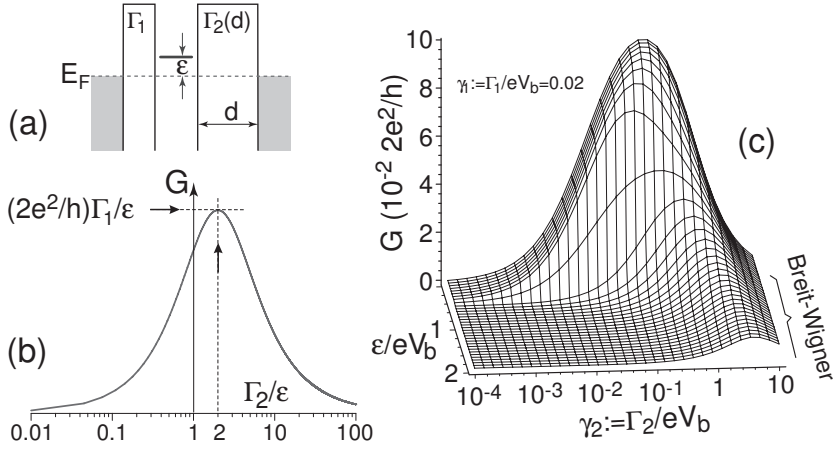}
\caption{\label{figure3}
    (a) Schematics of the energetics of a double-barrier junction with
    a single molecular level positioned at $\varepsilon$ relative to the
    Fermi energy $E_F$ of the metallic reservoirs.
    (b) Linear-response conductance $G$ as a function of $\Gamma_2$ for
    the double-barrier shown in (a), assuming coherent tunneling and
    $\Gamma_1 \ll \varepsilon$.
    (c) Same as (b), but taking into account the finite bias voltage
    $V_b \gg k_b T$. This illustration has been obtained for a specific value
    of \mbox{$\Gamma_1=0.02$\,$V_b$}.
    A pronounced crossover is observed at $\varepsilon = eV_b/2$.
    }
\end{figure}

In the first step, we obtain $\kappa$ by focussing on large $d$ where
the junction is widely open and the conductance well below the
conductance at the peak. In this simple tunneling regime,
$G\propto\Gamma_{2}\propto e^{-\kappa d}$. From fits to the averaged
conductance curves of the C$_{60}$-modified junctions (Fig.~\ref{figure2} insets),
we extract the inverse decay lengths
$\kappa_{DMSO} = 0.43 \pm 0.01$\,\AA$^{-1}$ and
$\kappa_{toluene} = 0.37 \pm 0.05$\,\AA$^{-1}$.
We emphasize that the value obtained for DMSO is more
reliable than the one for toluene, since the fit could be performed over a
more extended range of conductance.
This is due to the presence of a lower
conductance peak in toluene (see Fig.~\ref{figure2}b).
The inverse decay lengths $\kappa$ are slightly suppressed as
compared to the values measured in the solvent alone \cite{grueter04}.

\begin{figure}
\includegraphics*[width=11cm]{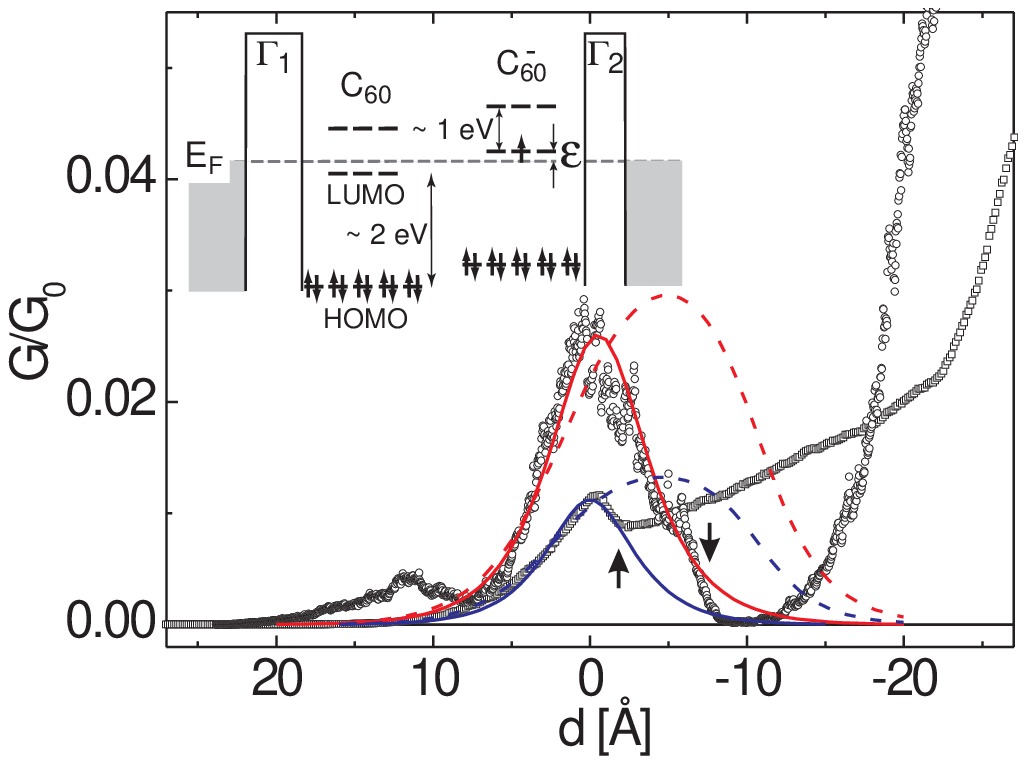}
\caption{\label{figure4}
    Averaged conductance versus distance curves for DMSO ({\tiny
    $\Box$}) and toluene ($\circ$).
    The solid lines are fits to the data using Eq.~(\ref{eq1})
    with $\Gamma_1/\varepsilon=1.1\cdot 10^{-2}$ ($2.7\cdot 10^{-2}$) and
    $\kappa=0.43$\AA$^{-1}$ ($0.37$\AA$^{-1}$) for DMSO and (toluene).
    The dashed lines were calculated from Eq.~(\ref{eq2})
    in the limit $\varepsilon=0$:
    \mbox{$\Gamma_{1}=0.92$\,meV} \mbox{($2.2$\,meV)},
    \mbox{$\Gamma_{2}^{\ast}=2.7$\,meV} \mbox{($3.9$\,meV)}, and
    \mbox{$\kappa=0.43$\,\AA$^{-1}$} \mbox{($0.37$\,\AA$^{-1}$)} for DMSO and (toluene).\\
    \emph{Inset}:
    Schematic energy diagram of
    the metal-molecule-metal junction showing the energy level
    of C$_{60}$ and C$_{60}^{-}$. Due to its large
    electronegativity, C$_{60}$ tends to gain charge from a nearby
    metallic surface as illustrated by the position of its LUMO level,
    lying below the Fermi energy of the Au electrodes.
    }
\end{figure}

We next compare our data to conductance curves calculated
in the limit $\varepsilon=0$. We fix the $\kappa$ values to the ones
determined for large $d$ and try to obtain a good match with the experiment
by tuning both $\Gamma_1$ and $\Gamma_2^{\ast}$.
The dashed lines in Fig.~\ref{figure4} show curves for
$\Gamma_{1,DMSO} = 0.92$\,meV and $\Gamma_{2,DMSO}^{\ast} = 2.7$\,meV, and
$\Gamma_{1,toluene} = 2.2$\,meV, $\Gamma_{2,toluene}^{\ast} = 3.9$\,meV.
In both cases, the adjusted curves present a
substantially broader peak than the data, showing that the limit
$\varepsilon=0$ is inappropriate.

We now relax the condition $\varepsilon = 0$ and try to find
the {\em smallest} possible $\varepsilon$ for which a good match
between the calculated and measured curve is obtained, again using
the $\kappa$ values determined for a large $d$. We obtain
$\varepsilon_{DMSO} \simeq 0.20$\,eV, $\Gamma_{1,DMSO} \simeq
4.3$\,meV, $\Gamma_{2,DMSO}^{\ast} \simeq 0.35$\,eV, and
$\varepsilon_{toluene} \simeq 0.12$\,eV, $\Gamma_{1,toluene}
\simeq 5.5$\,meV, $\Gamma_{2,toluene}^{\ast} \simeq 74$\,meV.
Care has to be taken at this point, because
\mbox{$\varepsilon_{toluene} \simeq 0.12$\,eV} is very close to
the transition region between the Breit-Wigner and the
$\varepsilon=0$ case where the theoretical $G(d)$ curves are quite
sensitive to the temperature. When taking the finite temperature
into account, $\varepsilon_{toluene}$ slightly increases, so that
\mbox{$\varepsilon \gtrsim 0.15-0.2$\,eV} is a good approximation
for both DMSO and toluene.

If we allow for larger values of $\varepsilon$ we cross over to
the Breit-Wigner equation. In this scale-invariant limit, we
obtain convincing fits for both DMSO and toluene with the
parameters $\Gamma_1/\varepsilon = 0.011$ (DMSO) and $0.027$
(toluene), and $\Gamma_2^{\ast}/\varepsilon=2$ as required (solid
lines in Fig.~\ref{figure4}). Hence, at this stage we can say with
confidence that \mbox{$\varepsilon \gtrsim 0.15-0.2$\,eV} for both
DMSO and toluene.

As emphasized before, the Breit-Wigner formula cannot provide an upper bound
for $\varepsilon$. However, we can -- on a physical ground -- provide an upper bound
on $\Gamma_2$ and therefore also on $\varepsilon$. The latter is obtained
through the relation $\Gamma_2^{\ast}=2\varepsilon$. The peak
in $G(\Gamma_2)$ appears at relatively large $\Gamma_2$ values, but
$\Gamma_2$ cannot grow to too high values, because otherwise {\em higher}
molecular states of the C$_{60}$ molecule will strongly hybridize with the states
in the Au electrode and the model will break down. Moreover, a large
$\Gamma_2$ at short distances corresponds to a large binding force
of order $\kappa\Gamma_2$ between the molecule and the Au electrode.
In fact, we think that the crossover from the peak to the plateau in
DMSO is determined by this consideration. Furthermore, the fully developed peak in
toluene suggests that $\Gamma_2$ at the position of the peak
($\Gamma_2^{\ast}$) is smaller in toluene than in DMSO.

In order to fix the upper bound of $\Gamma_2$ we need to discuss the
typical energy scales of the orbitals in C$_{60}$ \cite{comment1}.
In the inset of Fig.~\ref{figure4} we show the energy levels of C$_{60}$
and C$_{60}^{-}$ (adapted from Green {\it et al.} \cite{green96}).
The HOMO-LUMO gap of C$_{60}$ determined in solution is
around \mbox{$2.3$\,eV} \cite{echegoyen98} whereas local density
approximation calculations provide a value of \mbox{$1.9$\,eV} for the free
C$_{60}$ molecule (see e.g. \cite{dresselhaus96_12} and refs. therein).
Experiments on C$_{60}$ monolayers assembled on metallic
surfaces showed values ranging from \mbox{$1.5$\,eV} to \mbox{$3.0$\,eV}
\cite{maxwell94, gensterblum91,wang99, Lu04}.

Due to its high electronegativity, a C$_{60}$ molecule in contact with a
metal tends to gain charge from the metal. The ionization potential of Au is
\mbox{$5.3$\,eV} and the electron affinity of C$_{60}$ equals \mbox{$2.8$\,eV}. The latter includes
the Coulomb energy of the singly charged C$_{60}$ amounting to \mbox{$\approx 3$\,eV}.
If we assume that the Coulomb energy is strongly screened in the gap
between the Au electrodes, C$_{60}\rightarrow$C$_{60}^{-}$ is favorable.
Photoemission studies of C$_{60}$ on a Au surface yield values for
the electron charge on C$_{60}$
of $0.8$ \cite{tzeng00} and $1.0$ \cite{bart98}.
Hence, it is very likely that C$_{60}$ is singly charged
also in our work. On acquiring an electron
the threefold degenerate LUMO state of C$_{60}$ splits  by a very small amount
of \mbox{$8$\,meV} (not shown and also not resolvable in our experiment)
and the chemical potential $E_F$ is expected to align closely with this
orbital (dashed line). In fact, $E_F$ may lie above ($\varepsilon > 0$) or
below ($\varepsilon < 0$), depending on the exact charge state.

Taking the HOMO-LUMO gap of the C$_{60}$ and the C$_{60}$ anion as the proper energy scales,
\mbox{$\Gamma_2 \lesssim 1\dots 2.3$\,eV} to avoid hybridization
with higher lying orbitals. If we impose that this condition
occurs in DMSO at the transition from the peak to the observed
plateau at \mbox{$d=-2$\,\AA} (see arrow in Fig.~\ref{figure4}),
we obtain \mbox{$\varepsilon_{DMSO}< 0.2 \dots 0.5$\,eV}. Together
with the previous consideration, $\varepsilon$ is determined quite
accurately to \mbox{$0.2 < \varepsilon_{DMSO} < 0.5$\,eV}.
Similarly, if we pretend to describe the full peak in toluene up
to \mbox{$d=-7$\,\AA} (arrow), we obtain
\mbox{$\varepsilon_{toluene}\lesssim 0.04 \dots 0.09$\,eV}. These
rather small values for $\varepsilon$ are inconsistent with the
previous consideration. In order to obtain an upper bound for
$\varepsilon_{toluene}$ of at least \mbox{$0.2$\,eV}, one would
have to set the cut-off to as large a value as \mbox{$\Gamma_2\leq
5$\,eV}. This disagreement shows that the full decay of $G(d)$ in
toluene to approximately zero conductance beyond the peak must
have another origin. This will be addressed in the following.

We have assumed here that we are allowed the use the resonant
tunneling model with an unperturbed level beyond the peak.
However, a full understanding of the small $d$ regime would require the
understanding of local rearrangement effects taking place when
the electrodes and the molecules are reaching mechanical contact.
In the case of toluene we observe that the conductance drops to
a negligibly small value after the peak and then shows a pronounced increase,
similarly to the behavior of the junction in the pure solvent
(Fig.~\ref{figure2}b). We can presume that, upon closing of the
junction to small values of $d$, the molecules are pushed away from the inter-electrode
space, resulting in conductance properties dominated by Au-Au
tunneling. The shoulder observed in the case of DMSO
(Fig.~\ref{figure2}a) lets envision a different scenario in
which the molecules remain within the junction and
therefore continue to substantially affect the
conductance of the junction at very short inter-electrode
separations. Additionally, large conductance fluctuations can be observed in
toluene (see Fig.~\ref{figure2}b) while the curves are much
smoother in DMSO. We attribute these differences to the solubility
of the fullerene derivative in the solvent considered. Whereas
toluene is a good solvent, the fullerenes tend to aggregate with
time when suspended in DMSO.
In our view, the large conductance
fluctuations observed in toluene reflect the tendency of the
fullerene molecules to be re-solved if the cavity is getting to narrow.
In contrast, the molecules rather stay within the junction
if DMSO is used due to the lower solvation energy.

Whether a single molecule picture really
holds in such a system is an important question.
The very good match between theory and experiment
supports the single molecule approach. The analysis
provides more than a qualitative agreement with the data and
allows the extraction of narrowed range estimates for the LUMO
position and the tunneling rate $\Gamma_1$ of the molecular bridge
between the Au electrode and the C$_{60}$ molecule.
The results are:
\mbox{$\varepsilon = 0.2\dots 0.5$\,eV},
and \mbox{$\Gamma_1/\varepsilon = 1.1\dots 2.7\cdot 10^{-2}$}.

The above discussion is based on a non-interacting model, where
the interaction energy has been absorbed in the level energy
$\varepsilon$. When the coupling $\Gamma_2$ to the molecule
changes, the average charge on the molecule is modified, thereby
changing the interaction contribution to $\varepsilon$. However,
we have numerically checked (within a self-consistent mean field
theory \cite{zahid03}) that this effect does not appreciably
change the shape of the conductance versus distance curves, but
only slightly scales the parameters $\varepsilon$, $\Gamma_1$ and
$\Gamma_2^{\ast}$.

The presence of a much lower peak in some raw conductance curves
in toluene (about \mbox{$12$\,\AA} prior to the main peak,
Fig.~\ref{figure2}b) might be a signature for a weaker tunneling
process involving C$_{60}$ molecules on both electrodes. If
running experiments confirm this feature, an interpretation in
terms of a triple-barrier system
could further support this approach.

\section{Conclusions}

We have performed transport measurements through C$_{60}$
molecules tethered by a single anchor group in Au break junctions
operated in a liquid environment. The signature of the presence of the molecules
within the junction appeared as a peak in conductance versus
distance curves. The shape of this peak was strongly influenced by
the solvent in which the junction was operated, thereby showing
the importance of a proper environmental control in nanoscale
junctions.

The data can be understood in a resonant tunneling picture through
a single level located close to the Fermi energy $E_F$. The
particular geometry of the experiment combined with the
environmental control allowed us to provide numbers for
the electronic tunneling rates between a gold electrode and a
fullerene derivative.
By systematically varying the linker group,
our approach will allow to better understand the
electronic coupling between molecules and electrodes in molecular
devices. Moreover, gating will allow the resonant level to be
moved even closer to the Fermi level. An increase of the conductance up to the
quantum conductance value can then be envisaged.

\begin{ack}

We acknowledge the help of P. Morf for reductive desorption
measurements. M.T.G. acknowledges a grant from the "Ministerio de
Educaci{\'o}n y Ciencia". This work benefitted from the support of
the Swiss National Center of Competence in Research ``Nanoscale
Science'', the Eurocores Programme on Self-Organized
Nano-Structures and the Swiss National Science Foundation.

\end{ack}

\section*{References}

\bibliographystyle{apsrev}

\begin{thebibliography}{99}
\bibitem{comment1}
  We do not take into account the modifications in the orbitals induced by the
  anchor group attached to C$_{60}$.
\end{thebibliography}


\newpage

\clearpage

\end{document}